\documentclass[aps,prd,twocolumn,nofootinbib,superscriptaddress,10pt,floatfix]{revtex4-2}

\usepackage{amsmath,amssymb}
\usepackage{bm}
\usepackage{slashed}
\usepackage{graphicx}
\usepackage{hyperref}
\begin{document}

\title{Unpolarized Generalized Parton Distributions of the Nucleon in a Covariant Quark--Diquark Approach}

\author{Hao Zhang}
\email{zhangh01@ihep.ac.cn}
\affiliation{Institute of High Energy Physics, Chinese Academy of Sciences, Beijing 100049, China}
\affiliation{School of Physical Sciences, University of Chinese Academy of Sciences, Beijing 101408, China}

\author{Dongyan Fu}
\email{dy.fu@qq.com}
\affiliation{College of Physics, Guizhou University, Guiyang 550025, China}
\affiliation{Southern Center for Nuclear-Science Theory (SCNT), Institute of Modern Physics, Chinese Academy of Sciences, Huizhou 516000, China}

\author{Jiaqi Wang}
\email{jqwang@impcas.ac.cn}
\affiliation{Institute of High Energy Physics, Chinese Academy of Sciences, Beijing 100049, China}
\affiliation{School of Physical Sciences, University of Chinese Academy of Sciences, Beijing 101408, China}
\affiliation{Quark Matter Research Center, Institute of Modern Physics, Chinese Academy of Sciences,
Lanzhou, 730000, China}

\author{Yubing Dong}
\email{dongyb@ihep.ac.cn}
\affiliation{Institute of High Energy Physics, Chinese Academy of Sciences, Beijing 100049, China}
\affiliation{School of Physical Sciences, University of Chinese Academy of Sciences, Beijing 101408, China}

\date{November 18, 2026}

\begin{abstract}

We study the unpolarized generalized parton distributions (GPDs) of the nucleon in a covariant quark--diquark approach, in which both scalar and axial-vector diquarks are treated as composite subsystems.
Motivated by previous form factor studies, we introduce a vertex correction based on the pion cloud picture as a first exploratory treatment of pion cloud effects at the GPD level, in order to improve the description of $E^q(\xi,\,x,\,t)$ and related magnetic observables.
At the model input scale, only valence quarks are included.
With the vertex correction, the electromagnetic form factors $G_{E,M}$ and the mass form factor $A$ obtained from the Mellin moments of the GPDs agree well with those obtained from direct calculations.
The vertex correction brings the evolved charge weighted distribution $\sum{e^q E^q(\xi,\,x,\,t)}$ closer to the phenomenological parametrization.
The correction shifts transverse strength from the central region toward the peripheral region, consistent with the long range character of the pion cloud contribution.
The transverse structure results further suggest a harder longitudinal momentum distribution in the central region of the nucleon than in its periphery.
Furthermore, the evolved real part of the proton Compton form factor, $\mathrm{Re}\,\mathcal{H}(\xi,\,t)$, has a reasonable magnitude and follows the main trend of the Quantum Deep Neural Networks (QDNN) extraction.
In addition, $\mathrm{Re}\,\mathcal{H}$ may be more sensitive to the ERBL region than the other physical quantities calculated in this work.

\end{abstract}

\maketitle

\section{Introduction}
\label{sec:intro}

The internal structure of the nucleon has long been a central theme in hadron physics and remains one of the clearest windows into the nonperturbative dynamics of quantum chromodynamics (QCD).
Traditionally, our knowledge of the nucleon structure has been assembled from two complementary sources.
On the one hand, electromagnetic form factors (EMFFs)~\cite{Perdrisat:2006hj}, extracted from elastic lepton--nucleon scattering experiments~\cite{Hofstadter:1955ae}, characterize the electromagnetic structure of the nucleon in momentum space.
In particular, they can be interpreted in terms of transverse charge and magnetization densities~\cite{Miller:2007uy,Carlson:2007xd,Burkardt:2000za} in the infinite momentum frame, providing spatial information on the nucleon's transverse structure.
On the other hand, parton distribution functions (PDFs), extracted from deep inelastic scattering (DIS)~\cite{Collins:1989gx,Gao:2017yyd}, describe the parton distributions as functions of the light cone momentum fraction.
The EMFFs and PDFs thus provide complementary projections of nucleon structure.
A more complete description requires a framework in which the information contained in PDFs and form factors is combined.

Generalized parton distributions (GPDs) provide such a framework~\cite{Muller:1994ses,Ji:1996ek,Radyushkin:1996nd}.
Moreover, PDFs are obtained from GPDs in the forward limit, and the zeroth moments of GPDs give EMFFs~\cite{Ji:1996ek,Radyushkin:1996nd,Diehl:2003ny}.
At zero skewness, they connect the light cone momentum fraction of a parton with its transverse position and give access to impact parameter distributions (IPDs)~\cite{Burkardt:2000za,Diehl:2003ny,Guidal:2013rya}.
GPDs also contain information beyond PDFs and EMFFs.
Their first Mellin moments are related to the form factors of the QCD energy-momentum tensor (EMT), usually called gravitational form factors (GFFs).
These moments connect GPDs with the Ji sum rule~\cite{Ji:1996ek,Ji:1997gm} for the partonic decomposition of the nucleon spin.
GFFs also provide access to the internal pressure and shear force distributions of the nucleon~\cite{Polyakov:2002yz,Polyakov:2018zvc,Neubelt:2019sou}.

Experimentally, GPDs are primarily accessed through hard exclusive processes, most notably deeply virtual Compton scattering (DVCS)~\cite{Ji:1996nm,Radyushkin:1996nd,Collins:1996fb,Belitsky:2001ns}.
In DVCS analyses, the measured cross sections and spin asymmetries are usually described in terms of the complex Compton form factors (CFFs)~\cite{Belitsky:2001ns,Belitsky:2005qn}, which provide the phenomenological link between GPDs and observables~\cite{Guidal:2013rya,diaz2025globaldeepneuralnetwork,Le:2025swl}.
DVCS measurements have been reported by the Jefferson Lab Hall A~\cite{JeffersonLabHallA:2006prd}, CLAS~\cite{CLAS:2001wjj,CLAS:2007clm}, HERMES~\cite{HERMES:2001bob}, and COMPASS~\cite{COMPASS:2018pup}.
Looking ahead, the Electron-Ion Collider (EIC) in the United States~\cite{Accardi:2012qut,AbdulKhalek:2021gbh} and the planned Electron Ion Collider in China (EicC) will significantly enlarge the accessible kinematic domain and further sharpen constraints on both GPDs and CFFs~\cite{Anderle:2021wcy}.

On the theoretical side, lattice QCD continues to make important progress on low moments of GPDs~\cite{Bali:2018zgl} and on quasi-distribution approaches to their $x$ dependence~\cite{Scapellato:2022mai,Constantinou:2020hdm}.
Light front and constituent quark descriptions~\cite{Scopetta:2002xq,Boffi:2007yc}, chiral quark soliton studies~\cite{Goeke:2001tz,Ossmann:2004bp}, diquark spectator calculations~\cite{Fu:2023dea}, and covariant Nambu--Jona-Lasinio (NJL) based Faddeev approaches with quark--diquark correlations~\cite{Freese:2019eww} provide complementary access to GPD structure.
Nevertheless, determining the full $(\xi,\,x,\,t)$ dependence from first principles remains difficult, so nonperturbative phenomenological models continue to play an essential role in building intuition, interpreting data, and providing realistic inputs for phenomenological studies.

The NJL based covariant quark--diquark study has shown that dynamical scalar and axial-vector diquark correlations strongly affect the flavor structure of the proton GPDs and the form factors obtained from the Mellin moments~\cite{Freese:2019eww}.
However, the usual diquark spectator treatment does not resolve the quarks inside the diquark~\cite{Fu:2023dea}.
To obtain a more complete description of the internal structure of the nucleon, we include the internal quark structure of the diquark in the nucleon GPDs.
Previous form factor calculations in the covariant quark--diquark framework showed that the pion cloud plays an important role in nucleon form factors~\cite{Wang:2024abv}.
Without this contribution, the magnetic form factors and the D-term are not well described.
Meson cloud effects have also been included in GPD studies~\cite{Pasquini:2006dv,Pasquini:2006ib}, and similar corrections have been applied to the EMFFs of the $\Omega^-$~\cite{Fu:2025vkq}.
These results motivate us to make an exploratory attempt to include a pion cloud motivated correction at the GPD level.

This paper is organized as follows.
In Sec.~\ref{sec:formalism} we summarize the formal definitions of the nucleon GPDs, local form factors, CFFs, and transverse structure.
Section~\ref{sec:model} introduces the covariant quark--diquark approach, the convolution formula and the vertex correction.
In Sec.~\ref{sec:results}, we present the calculated GPDs and compare the corresponding form factors with direct calculations, examine the changes induced by the vertex correction, and discuss further results with the correction.
Finally, Sec.~\ref{sec:summary} contains the summary and outlook.

\section{Formal Definitions}
\label{sec:formalism}

\subsection{Kinematics and GPD Definitions}
For the nucleon, a spin-$1/2$ particle, the unpolarized GPDs $H^q(\xi,\,x,\,t)$ and $E^q(\xi,\,x,\,t)$ are defined through the nonforward matrix element of the twist-2 light cone quark correlator~\cite{Ji:1996ek,Radyushkin:1996nd,Diehl:2003ny,Mondal:2015uha},
\begin{widetext}
\begin{align}
  \mathcal{M}_{\lambda'\lambda}^q(\xi,\,x,\,t)
  &\equiv \frac{1}{2} \int \frac{dz^-}{2\pi} e^{i x (P\cdot z)}
  \left\langle p',\,\lambda' \left|
  \bar{\psi}_q \left(-\frac{z}{2}\right) \slashed{n}
  \psi_q \left(\frac{z}{2}\right)
  \right| p,\,\lambda \right\rangle \Big|_{z^+=\bm{z}_\perp=0}
  \nonumber \\
  & =
  \frac{1}{P^+}\bar{u}(p',\,\lambda')
  \left[
    H^q(\xi,\,x,\,t) \slashed{n}
    + E^q(\xi,\,x,\,t) \frac{i \sigma^{nq}}{2M}
  \right]
  u(p,\,\lambda),
  \label{eq:GPD_def}
\end{align}
\end{widetext}
where $p$ and $p'$ respectively denote the incoming and outgoing nucleon momenta, $P=(p+p')/2$ is the average nucleon momentum, $q=p'-p$ is the momentum transfer, and $t=q^2$.
From the definitions, one obtains the frequently used relation $P \cdot q = 0$ because the initial and final particles have the same mass.
The variable $x$ denotes the light cone momentum fraction carried by the parton and is defined explicitly in Sec.~\ref{sec:model}.
The skewness parameter is $\xi=-\frac{q^+}{2P^+}$.
In the numerical calculation, we take $\xi \leq 0$.
In addition, $M$ denotes the nucleon mass, and the Dirac spinors are normalized as $\bar{u}(p,\,\lambda')u(p,\,\lambda)=2M\delta_{\lambda'\lambda}$.
In this work, we use light cone coordinates $u^\pm=u^0\pm u^3$, so that a four-vector is written as $u^\mu=(u^+,u^-,\bm{u}_\perp)$ and the scalar product is $u\cdot v=\frac{1}{2}u^+v^-+\frac{1}{2}u^-v^+-\bm{u}_\perp\cdot\bm{v}_\perp$.
The lightlike vector $n^\mu = (0,\,2,\,\bm{0}_\perp)$ is chosen to project out the plus component, $u\cdot n=u^+$, and we use the convention $i\sigma^{nq}=i\sigma^{\mu\nu}n_\mu q_\nu$.

\subsection{Local Form Factors and Sum Rules}

The Mellin moments of GPDs are polynomials in $\xi$, with coefficients related to local form factors.
This polynomiality structure follows directly from Lorentz covariance~\cite{Ji:1996ek,Diehl:2003ny,Belitsky:2005qn}.

\subsubsection{Electromagnetic Form Factors}

The matrix element of the local electromagnetic current is usually decomposed as~\cite{Sachs:1962zzc,Wang:2024abv}
\begin{align}
  &\langle p',\,\lambda'|J_q^\mu|p,\,\lambda\rangle \nonumber \\
  =&\,
  \bar{u}(p',\,\lambda')
  \left[
    \gamma^\mu F_1^q(t)
    +\frac{i\sigma^{\mu q}}{2M}F_2^q(t)
  \right] u(p,\,\lambda),
  \label{eq:em_current_decomp}
\end{align}
with $J_q^\mu=\bar{\psi}_q(0)\gamma^\mu \psi_q(0)$.
Projecting this local current onto the light cone direction and comparing with the $x$ integral of Eq.~\eqref{eq:GPD_def}, we obtain the following sum rules,
\begin{equation}
  \begin{aligned}
  \int_{-1}^1 dx\,H^q(\xi,\,x,\,t)
  &= F_1^q(t), \\
  \int_{-1}^1 dx\,E^q(\xi,\,x,\,t)
  &= F_2^q(t).
  \end{aligned}
  \label{eq:zero sum rules}
\end{equation}

The total nucleon Dirac and Pauli form factors are then written as $F_{1,2}(t)=\sum_q e_q F_{1,2}^q(t)$, where $e_q$ is the electric charge of quark flavor $q$.
Their linear combinations produce the Sachs electric and magnetic form factors~\cite{Sachs:1962zzc,Ernst:1960zza},
\begin{equation}
  \begin{aligned}
    G_E(t) &= F_1(t) + \frac{t}{4M^2} F_2(t), \\
    G_M(t) &= F_1(t) + F_2(t).
  \end{aligned}
\end{equation}
One obtains the charge $Q_e=G_E(0)$ and magnetic moment $\mu=\frac{e}{2M}G_M(0)$ when the momentum transfer squared $t\to 0$.

\subsubsection{Gravitational Form Factors}

The nucleon GFFs are extracted from the matrix element of the quark EMT~\cite{Pagels:1966zza,Wang:2024abv},
\begin{equation}
\begin{split}
  &\langle p',\,\lambda'|T^{\mu\nu}_q|p,\,\lambda\rangle \\
  = &
  \bar{u}(p',\,\lambda')
  \left[
    \frac{P^\mu P^\nu}{M}F^q_{1,0}(t) + \frac{q^\mu q^\nu-g^{\mu\nu}q^2}{M}F^q_{2,0}(t) \right. \\
   & \left. + M g^{\mu\nu}F^q_{3,0}(t) + \frac{i}{4M}P^{\{\mu}\sigma^{\nu\}q}F^q_{4,0}(t)
  \right]
  u(p,\,\lambda),
\end{split}
\label{eq:emt_current_decomp}
\end{equation}
where $X^{\{\mu}Y^{\nu\}}=X^\mu Y^\nu+X^\nu Y^\mu$ denotes symmetrization over the Lorentz indices.
Here $T_q^{\mu\nu}=\frac{i}{2}\bar{\psi}_q\gamma^{\{\mu}\overleftrightarrow{\partial}^{\nu\}}\psi_q$ is the symmetric quark EMT used in the present model~\cite{Fu:2022rkn}.
The operator $\overleftrightarrow{\partial}^{\mu}=\frac{1}{2}(\overrightarrow{\partial}^{\mu}-\overleftarrow{\partial}^{\mu})$ is half the difference between the derivatives acting on the emitted and absorbed quark fields.

The functions $F^q_{i,0}$ defined in Eq.~\eqref{eq:emt_current_decomp} are related to the physical form factors through the following linear combinations~\cite{Wang:2024abv,Ji:1997gm,Diehl:2003ny},
\begin{equation}
\begin{aligned}
  A^q(t) &= F^q_{1,0}(t), \\
  B^q(t) &= F^q_{4,0}(t)-F^q_{1,0}(t), \\
  J^q(t) &=\frac{1}{2}\left[A^q(t)+B^q(t)\right]
  =\frac{1}{2}F^q_{4,0}(t), \\
  D^q(t) &= 4F^q_{2,0}(t).
\end{aligned}
\label{eq:angular_form_factor}
\end{equation}
Summing over quark flavors gives the nucleon GFFs, $A(t)=\sum_q A^q(t)$, $B(t)=\sum_q B^q(t)$, $J(t)=\sum_q J^q(t)$ and $D(t)=\sum_q D^q(t)$, where $A(t)$, $B(t)$, $J(t)$, and $D(t)$ are the mass or energy~\cite{Belitsky:2005qn,Polyakov:2018zvc}, gravitomagnetic~\cite{Belitsky:2005qn,Polyakov:2018zvc,Freese:2019eww}, angular momentum, and D-term~\cite{Polyakov:2002yz,Polyakov:2018zvc,Neubelt:2019sou} form factors, respectively.
Moreover, the form factor $F^q_{3,0}(t)$ corresponds to the $g^{\mu\nu}$ term in this decomposition and does not enter the first Mellin moments of the leading-twist unpolarized GPDs after the light cone projection since $n_\mu n_\nu g^{\mu\nu} = 0$.
The GFFs can also be obtained by taking the first Mellin moments of the unpolarized GPDs~\cite{Ji:1997gm,Diehl:2003ny,Polyakov:2002yz,Wang:2024abv},
\begin{equation}
  \begin{aligned}
    &\int_{-1}^1 dx\,x H^q(\xi,\,x,\,t)
    = A^q(t)+\xi^2D^q(t), \\
    &\int_{-1}^1 dx\,x E^q(\xi,\,x,\,t)
    = B^q(t)-\xi^2D^q(t),\\
    &\frac{1}{2}\int_{-1}^1 dx\,x \left(H^q(\xi,\,x,\,t)+E^q(\xi,\,x,\,t)\right)
    = J^q(t).
  \end{aligned}
  \label{eq:first sum rules}
\end{equation}

\subsection{Compton Form Factors}
Within the factorization framework for DVCS, the DVCS amplitude is given by the convolution of the GPDs with the perturbative coefficient function~\cite{Collins:1996fb,Belitsky:2001ns,Belitsky:2005qn}.
For the generic unpolarized GPDs $F=H,E$, the corresponding CFFs $\mathcal{F}=\mathcal{H},\mathcal{E}$ may be written as~\cite{Belitsky:2001ns,Belitsky:2005qn,Le:2025swl}
\begin{equation}
  \mathcal{F}(\xi,\,t)
  = \sum_q e_q^2 \int_{-1}^{1} dx\,
  C(\xi,\,x)\,F^q(\xi,\,x,\,t).
  \label{eq:cff_definition}
\end{equation}
These form factors are complex quantities, and the leading order (LO) coefficient function is
\begin{equation}
  C(\xi,\,x)=\frac{1}{\xi-x-i\epsilon}-\frac{1}{\xi+x-i\epsilon}.
  \label{eq:cff_C_definition}
\end{equation}
The imaginary parts of the CFFs are determined by the GPD values on the crossover lines $x=\pm|\xi|$, whereas the real parts are given by principal value integrals.
Thus, the CFFs provide the connection between the GPDs and DVCS observables and are sensitive to both the local behavior of the GPDs at the DGLAP--ERBL boundaries and their global structure in $x$.

\subsection{Normalized Second Transverse Moment and Transverse Charge Densities}

At zero skewness, a two-dimensional Fourier transform with respect to the transverse momentum transfer $\bm{q}_\perp$ yields the IPDs~\cite{Miller:2007uy,Burkardt:2000za,Miller:2010nz},
\begin{align}
  q_N(x,\,\bm{b}_\perp) =
  \int \frac{d^2\bm{q}_\perp}{(2\pi)^2}
  e^{-i\bm{b}_\perp\cdot\bm{q}_\perp}
  H_N(0,\,x,\,-\bm{q}_\perp^2),
  \label{eq:ipd_def}
\end{align}
where $H_N=\sum_q e_q H_N^q$.
We use the subscript $N=p,n$ to distinguish the proton and neutron GPDs.
This distribution can be interpreted as the charge weighted transverse distribution of partons with the light cone momentum fraction $x$.

Using azimuthal symmetry, Eq.~\eqref{eq:ipd_def} can be recast as the Hankel transform,
\begin{equation}
  q_N(x,\,b)=\frac{1}{2\pi}\int_0^\infty dQ\,Q\,
  J_0(bQ)\,H_{N}(0,\,x,\,-Q^2),
  \label{eq:ipd_hankel_formal}
\end{equation}
with $b=|\bm{b}_\perp|$ and $Q^2=\bm{q}_\perp^2=-t$.

After integrating over $x$, the IPDs reduce to the transverse charge densities of the nucleon,
\begin{equation}
\begin{aligned}
  \rho_N(b)
  &=\int_{-1}^1 dx\,q_N(x,\,b) \\
  &=\frac{1}{2\pi}\int_0^\infty dQ\,Q\,
  J_0(bQ)\,F_1^N(-Q^2).
\end{aligned}
  \label{eq:transverse_charge_density}
\end{equation}
These densities describe the distribution of parton charge at a transverse distance $b$ from the nucleon's transverse center of momentum~\cite{Miller:2007uy}.

For the nucleon, the normalized second transverse moment is defined as~\cite{Diehl:2004cx}
\begin{equation}
\begin{aligned}
  \langle \bm b^2\rangle_{x,N}
  &=\frac{\int_0^\infty db\,b^3\,q_N(x,\,b)}
          {\int_0^\infty db\,b\,q_N(x,\,b)} \\
  &=\left.\frac{4}{H_N(0,\,x,\,0)}
  \frac{\partial H_N(0,\,x,\,t)}{\partial t}\right|_{t=0},
\end{aligned}
  \label{eq:ipd_b2_def}
\end{equation}
where $N=p,n$ and the definition applies when $H_N(0,\,x,\,0)\neq 0$.
It characterizes the charge weighted mean squared transverse distance from the nucleon's transverse center of momentum at fixed $x$.

\section{Theoretical Framework}
\label{sec:model}

In the present work, the nucleon is described within a covariant quark--diquark framework as a bound state of a quark and a correlated diquark~\cite{Kroll:1991ag,Keiner:1995bu,Fu:2023dea}.
For the spin-$1/2$ nucleon, the diquark may be either scalar (spin-0) or axial-vector (spin-1), and its internal quark structure is included through the diquark GPDs.

\subsection{Covariant Quark--Diquark Approach}

Note that GPDs and the corresponding form factors can be evaluated in different frames~\cite{Belitsky:2005qn}.
In the present work, we employ the Breit frame.
The initial and final nucleon momenta are written explicitly as~\cite{Fu:2023dea}
\begin{equation}
  \begin{split}
    p &= \left(P^0-\frac{q_z}{2},\, P^0+\frac{q_z}{2},\, -\frac{\bm{q}_\perp}{2}\right), \\
    p' &= \left(P^0+\frac{q_z}{2},\, P^0-\frac{q_z}{2},\, \frac{\bm{q}_\perp}{2}\right),
  \end{split}
\end{equation}
which gives $q^+=-q^-$.

Assuming SU(6) spin-flavor symmetry, the proton and neutron flavor wave functions may be written as~\cite{Ma:2002ir}
\begin{equation}
  \begin{aligned}
    \lvert p \rangle
    &= \cos\theta \lvert u(ud)_s \rangle \\
    &\quad
    + \sin\theta \left[
    \sqrt{\frac{2}{3}} \lvert d(uu)_a \rangle
    - \sqrt{\frac{1}{3}} \lvert u(ud)_a \rangle
    \right], \\
    \lvert n \rangle
    &= \cos\theta \lvert d(ud)_s \rangle \\
    &\quad
    + \sin\theta \left[
    -\sqrt{\frac{2}{3}} \lvert u(dd)_a \rangle
    + \sqrt{\frac{1}{3}} \lvert d(ud)_a \rangle
    \right],
  \end{aligned}
  \label{eq:flavor_wf}
\end{equation}
where $(q_1q_2)_s$ and $(q_1q_2)_a$ respectively denote the scalar and axial-vector diquarks and $\theta$ is the mixing angle between the different diquark types. To simplify the model, we use the same mixing angle as in Ref.~\cite{Wang:2024abv} in this work.

\begin{figure*}[!t]
\centering
\includegraphics[width=2\columnwidth]{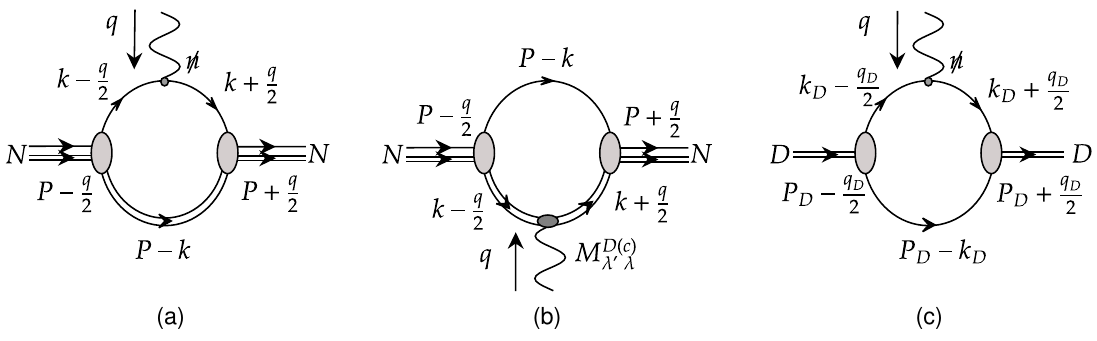}
\caption{\footnotesize{Feynman diagrams of the nucleon GPDs in the quark--diquark framework.
Panels (a) and (b) represent the quark and diquark distributions, respectively.
In panel (b), $\mathcal{M}_{\lambda'\,\lambda}^{D(c)}$ denotes the effective interaction vertex calculated from panel (c), which represents the quark level GPDs inside the diquark.}}
\label{fig:feynman}
\end{figure*}
Following the general strategy in Refs.~\cite{Fu:2023dea,Wang:2024abv}, we explicitly present the Feynman diagrams in Fig.~\ref{fig:feynman}.
With the momentum convention in Fig.~\ref{fig:feynman}, the emitted and absorbed partons carry momentum fractions $x+\xi$ and $x-\xi$, respectively, where $x=\frac{k^+}{P^+}$.
Then, for the axial-vector diquark channel in panel (a) of Fig.~\ref{fig:feynman}, the active quark GPDs are given by the amplitude~\cite{Fu:2023dea,Wang:2024abv,Fu:2022rkn},
\begin{widetext}
\begin{align}
  \mathcal{M}_{\lambda'\,\lambda}^{q}(\xi,\,x,\,t) &= -\frac{i}{2} c_1^2
  \int \frac{d^2 \bm{k}_{\perp}\,dk^+\,dk^-}{(2\pi)^4}
  \frac{\delta(k^+ - xP^+)}{\mathfrak{D}}\bar{u}(p') \Gamma_a^{\alpha}\left(\slashed{k}+\frac{\slashed{q}}{2}+m_q\right)
  \slashed{n}
  \left(\slashed{k}-\frac{\slashed{q}}{2}+m_q\right)g_{\alpha\beta}
  \Gamma_a^{\beta} u(p),
  \label{eq:qd_amplitude_a}
\end{align}
\end{widetext}
where the factor $g^{\alpha\beta}$ comes from the numerator of the axial-vector diquark propagator.
Compared with the complete Proca propagator, our propagator omits the longitudinal term $k^\alpha k^\beta/m_D^2$, where $m_D$ is the diquark mass.
Previous calculations found that this term gives only a small numerical contribution but considerably complicates the loop calculation~\cite{Fu:2023dea,Wang:2024abv}.
For the scalar diquark channel, $\Gamma_a^{\alpha(\beta)}$ is replaced by $\Gamma_s$, and $g_{\alpha \beta}$ is omitted because the scalar propagator has no tensor numerator.
Here $k$ denotes the average loop momentum flowing through the active quark line, and $m_q$ is the constituent quark mass.
For the scalar and axial-vector diquarks, the effective quark-nucleon-diquark vertices are taken as $\Gamma_s=c_s$ and $\Gamma_a^\alpha=c_a\gamma^\alpha\gamma^5$~\cite{Wang:2024abv,Scadron:1968zz}.
To simulate the bound state of the nucleon, an extra scalar function is attached to the vertex~\cite{Wang:2024abv},
\begin{equation}
  \Xi(p_1,\,p_2)=
  \frac{c_1}{(p_1^2-m_R^2+i\epsilon)(p_2^2-m_R^2+i\epsilon)},
  \label{eq:scalar_function}
\end{equation}
where $p_1$ ($p_2$) denotes the quark (diquark) momentum, $c_1$ is a coefficient in units of $\mathrm{GeV}^4$ and $m_R$ is the cutoff mass.
The denominator $\mathfrak{D}$ collects all the denominators from the propagators and the scalar functions~\cite{Fu:2023dea,Wang:2024abv},
\begin{widetext}
\begin{align}
  \mathfrak{D} &=
  \left[ \left(k + \frac{q}{2}\right)^{2} - m_{q}^{2} + i\epsilon \right]
  \left[ \left(k - \frac{q}{2}\right)^{2} - m_{q}^{2} + i\epsilon \right]
  \left[ (k - P)^{2} - m_{D}^{2} + i\epsilon \right] \nonumber \\
  &\quad \times
  \left[ (k - P)^{2} - m_{R}^{2} + i\epsilon \right]^{2}
  \left[ \left(k + \frac{q}{2}\right)^{2} - m_{R}^{2} + i\epsilon \right]
  \left[ \left(k - \frac{q}{2}\right)^{2} - m_{R}^{2} + i\epsilon \right].
  \label{eq:denominator}
\end{align}
\end{widetext}
The normalization factors $c_1c_s$ and $c_1c_a$ are fixed by the proton charge conditions $G_E^{p,s(a)}(0)=1$ in the scalar and axial-vector diquark channels, respectively. The cutoff mass $m_R$ in the vertex function is chosen close to the nucleon mass. Since the previous calculation showed a weak dependence of the EMFFs on $m_R$, we adopt the same value $m_R=1.0\,\mathrm{GeV}$ as in Ref.~\cite{Wang:2024abv}.
The internal quark structure of the diquark in Fig.~\ref{fig:feynman}(c) is treated following Ref.~\cite{Wang:2024abv}.
The scalar and axial-vector quark--quark--diquark vertices have the Dirac structures $\gamma^5$ and $\gamma^\alpha$, respectively.

\subsection{Convolution Formula}
\label{sec:convolution_formula}

In QCD, convolution descriptions are commonly motivated by the operator product expansion and renormalization group arguments~\cite{Collins:1989gx,Scopetta:2003et}.
A comparably rigorous derivation is usually not available in low energy effective theories.
Nevertheless, convolution models have long been used for composite hadronic systems and are effective~\cite{Jaffe:1983fw,Bickerstaff:1989ch}.
We therefore use the following convolution expression as the starting point,
\begin{equation}
  h(x)=\int \frac{dy}{y}\, f(y)\,
  g\left(\frac{x}{y}\right).
  \label{eq:xi0_convolution}
\end{equation}
In Eq.~\eqref{eq:xi0_convolution}, $h(x)$ denotes the parton distribution in the full system.
The function $f(y)$ is the distribution of the composite constituent in the parent system, where $y$ is the light cone momentum fraction carried by this constituent.
The function $g(x/y)$ describes the parton distribution inside the composite constituent, where $x/y$ is the internal light cone momentum fraction carried by the parton in the composite constituent.

\begin{figure*}[!t]
\centering
\includegraphics[width=2\columnwidth]{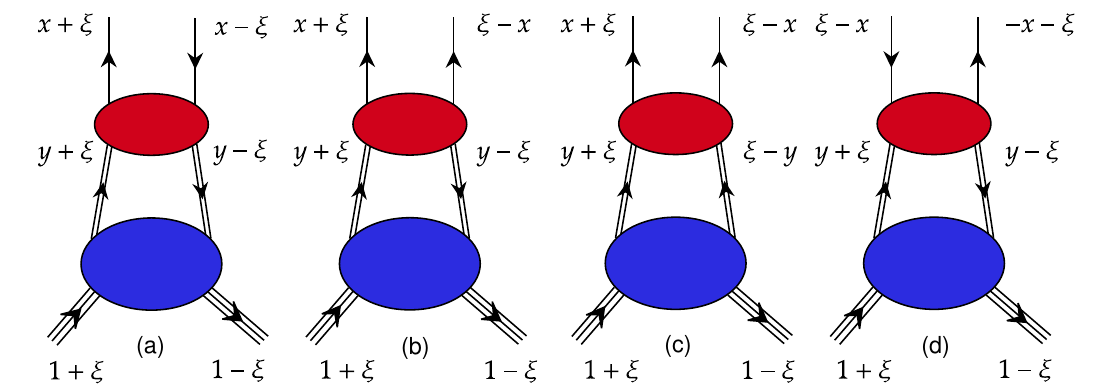}
\caption{\footnotesize{Schematic illustration of the different kinematic region combinations associated with the full convolution formula of Eq.~\eqref{eq:full_convolution}.}}
\label{fig:convolution_regions}
\end{figure*}

For GPDs, Eq.~\eqref{eq:xi0_convolution} must be extended to nonforward kinematics, where the internal variables become $x/y$ and $\xi/y$.
Following Ref.~\cite{Gao:2024ajz}, the DGLAP--ERBL separation into the regions $|x|\geq|\xi|$ and $|x|<|\xi|$~\cite{Muller:1994ses,Radyushkin:1996nd,Ji:1996ek,Diehl:2003ny,Belitsky:2005qn} is then applied to both the nucleon-level distribution and the internal diquark distribution, as illustrated in Fig.~\ref{fig:convolution_regions}.
The kinematic region assignments for the four diagrams are summarized in Table~\ref{tab:convolution_regions}.
\begin{table}[!htbp]
\caption{\footnotesize{Kinematic region assignments for the convolution diagrams in Fig.~\ref{fig:convolution_regions}. Here $q/N$, $D/N$, and $q/D$ denote the active quark in the nucleon, the diquark in the nucleon, and the quark in the diquark, respectively.}}
\label{tab:convolution_regions}
\begin{ruledtabular}
\begin{tabular}{l@{\quad}|@{\quad}ccc}
Panel & $q/N$ & $D/N$ & $q/D$ \\
\hline
(a) & $|\xi|<x<1$ & $|\xi|<y<1$ & $x/y>|\xi/y|$ \\
(b) & $|x|<|\xi|$ & $|\xi|<y<1$ & $|x/y|<|\xi/y|$ \\
(c) & $|x|<|\xi|$ & $|y|<|\xi|$ & GDA region \\
(d) & $x<-|\xi|$ & $|\xi|<y<1$ & $x/y<-|\xi/y|$
\end{tabular}
\end{ruledtabular}
\end{table}
Panel (c) represents a crossed channel generalized distribution amplitude (GDA) contribution, which is confined to the central region with $|x|<|\xi|$ and $|y|<|\xi|$.
Panel (d) is interpreted as the antiquark contribution from the diquark.

This organization leads to the full convolution formula,
\begin{widetext}
\begin{equation}
  H_D^{q}(\xi,\,x,\,t) =
  \begin{cases}
    \displaystyle\int_{x}^{1} \frac{dy}{y}\, f_D(\xi,\,y,\,t)\, H_{I}^{q}\!\left(\frac{\xi}{y},\,\frac{x}{y},\,t\right), & x>|\xi|, \\[2.5ex]
    \displaystyle\int_{|\xi|}^{1} \frac{dy}{y}\, f_D(\xi,\,y,\,t)\, H_{I}^{q}\!\left(\frac{\xi}{y},\,\frac{x}{y},\,t\right)\,+\,\int_{-|\xi|}^{|\xi|} \frac{dy}{2\xi}\, f_D(\xi,\,y,\,t)\, \frac{\xi}{\pi y} \int_{s_{0}}^{\infty} ds\, \frac{\mathrm{Im}\,\Phi\!\left(\frac{\frac{x}{\xi}+1}{2},\,\frac{\frac{y}{\xi}+1}{2},\,s\right)}{s-t+i\epsilon}, &|x|<|\xi|, \\[2.5ex]
    \displaystyle\int_{-x}^{1} \frac{dy}{y}\, f_D(\xi,\,y,\,t)\, H_{I}^{q}\!\left(\frac{\xi}{y},\,\frac{x}{y},\,t\right), & x<-|\xi|,
  \end{cases}
  \label{eq:full_convolution}
\end{equation}
\end{widetext}
where $f_D(\xi,\,y,\,t)$ denotes the GPD of the diquark in the nucleon, $H_I^q$ denotes the quark GPD inside the diquark, $\Phi$ is the diquark GDA, $s$ is the timelike squared invariant mass entering the GDA spectral representation in the crossed channel, and $s_0$ denotes the lowest physical threshold of this channel.
In the ERBL case $|x|<|\xi|$, the first term corresponds to panel (b) of Fig.~\ref{fig:convolution_regions}, while the second term is the crossed channel GDA contribution represented by panel (c).

In the numerical calculation, we keep the $x>|\xi|$ DGLAP term and the first term in the ERBL case $|x|<|\xi|$.
The $x<-|\xi|$ DGLAP term corresponds to an antiquark contribution and is absent in the present low scale valence input.
The second term in the ERBL case is the crossed channel GDA contribution and is left for future work.

\subsection{Vertex Correction}
\label{sec:vertex_correction}

A complete pion cloud contribution to GPDs should enter through the quark degrees of freedom associated with the pion cloud and then be embedded into the nucleon GPDs.
Such a calculation is technically involved.
In the present work, we therefore make an exploratory approximation and focus on the possible impact of the pion cloud.
We adopt a phenomenological implementation in which the long range effect suggested by the pion cloud picture is incorporated through an effective vertex correction.

\begin{figure*}[!t]
\centering
\includegraphics[width=2\columnwidth]{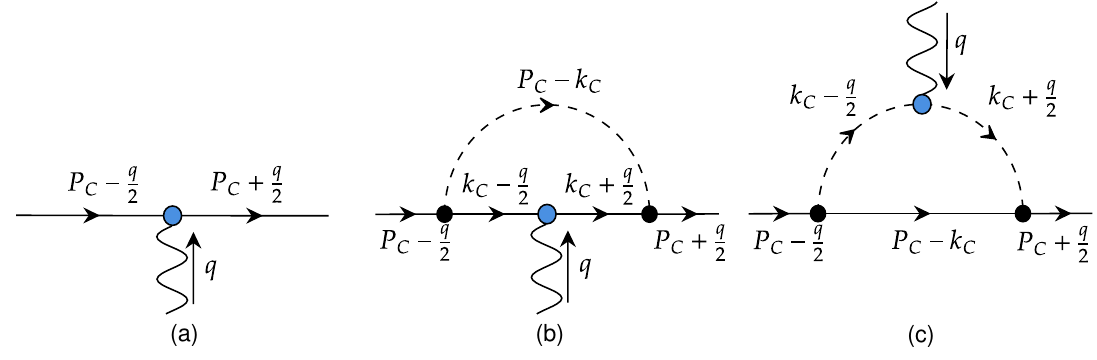}
\caption{\footnotesize{Coupling of the electromagnetic current to the quark and pion components in the presence of pion cloud corrections.
Panel (a) shows the coupling without the pion cloud, panel (b) the coupling of the photon to the quark in the presence of the cloud, and panel (c) the coupling of the photon to the pion cloud.}}
\label{fig:meson_cloud_diagrams}
\end{figure*}

Following the electromagnetic current parametrization for a spin-$1/2$ particle in Ref.~\cite{Wang:2024abv}, the effective light cone quark current vertex may be written as
\begin{equation}
    \Lambda^q(t)=\slashed{n}F_1^{q,\mathrm{VC}}(t)
  +\frac{i\sigma^{n q}}{2m_q}F_{2}^{q,\mathrm{VC}}(t).
  \label{eq:dressed_vertex}
\end{equation}
Here $\mathrm{VC}$ denotes vertex correction.
The form factors $F_i^{q,\mathrm{VC}}(t)$ are normalized such that $F_1^{q,\mathrm{VC}}(0)=1$.
After including the one loop pion--quark diagrams in Fig.~\ref{fig:meson_cloud_diagrams}, the dressed form factors in the electromagnetic current channel take the form
\begin{equation}
  F_i^{q,\mathrm{VC}}(t)=ZF_i^{q,PL}(t)
  +(1-Z)\left[f_i^{q}(t)+f_i^{\pi}(t)\right],
  \label{eq:meson_cloud}
\end{equation}
where $i=1,2$.
Here $F_i^{q,PL}$ denotes the bare core contribution, while $f_{i}^{q}$ and $f_{i}^{\pi}$ are generated by the quark coupling and pion coupling parts of the pion--quark loop, respectively.
The renormalization factor $Z$ stands for the probability of striking a quark without the pion cloud~\cite{Wang:2024abv}.
In Fig.~\ref{fig:meson_cloud_diagrams},  $P_C$ and $k_C$ denote the average momentum of the dressed quark and the loop momentum, respectively.

To apply this correction to GPDs, we need to specify how the dressed current insertion enters the moments of the light cone correlator.
The derivation of the standard Mellin moments is well known~\cite{Diehl:2003ny}\footnote{The Mellin moment expression in Ref.~\cite{Fu:2022bpf} does not account for the difference between its light cone coordinate convention and that of the cited reference. The associated overall factor is absorbed by the normalization, leaving the normalized numerical results unchanged.},
\begin{widetext}
\begin{equation}
  \begin{aligned}
  &\frac{(P^+)^{a+1}}{2}
  \int dx\,x^a \int\frac{dz^-}{2\pi} e^{ix(P\cdot z)}
  \left[
    \bar{\psi}_q\left(-\frac{z}{2}\right)\slashed{n}
    \psi_q\left(\frac{z}{2}\right)
  \right]\Big|_{z^+=\bm{z}_\perp=0}
  \\
  =&\,\left(2i\frac{d}{dz^-}\right)^a
  \left[
    \bar{\psi}_q\left(-\frac{z}{2}\right)\slashed{n}
    \psi_q\left(\frac{z}{2}\right)
  \right]\Big|_{z=0}
  =\,\bar{\psi}_q(0)\slashed{n}
  \left(i\overleftrightarrow{\partial}^{+}\right)^a
  \psi_q(0).
  \end{aligned}
\end{equation}
\end{widetext}
Here $a=0,1,2,\ldots$ denotes the order of the Mellin moment.
This identity allows us to implement the vertex correction at the model level by replacing the bare current insertion with the dressed one in Eq.~\eqref{eq:dressed_vertex}.

For the lowest moment, this gives
\begin{equation}
\begin{aligned}
\bar{\psi}_q(0)\left(\gamma^\mu F_1^{q,\mathrm{VC}}(t)
  +\frac{i\sigma^{\mu q}}{2m_q}F_{2}^{q,\mathrm{VC}}(t) \right)\psi_q(0),
\end{aligned}
\end{equation}
which corresponds to the electromagnetic pion cloud effect correction discussed in Ref.~\cite{Cloet:2014rja}.
Using this corrected vertex, we calculate the GPDs and extract the nucleon magnetic moments using Eq.~\eqref{eq:zero sum rules}.
The resulting magnetic moments are closer to the experimental values than those obtained without the vertex correction.

For the next moment, the same prescription yields
\begin{equation}
\begin{aligned}
\bar{\psi}_q(0) \left(\gamma^\mu F_1^{q,\mathrm{VC}}(t)
  +\frac{i\sigma^{\mu q}}{2m_q}F_{2}^{q,\mathrm{VC}}(t) \right)P_C^\nu \psi_q(0).
\end{aligned}
\end{equation}

Projecting the effective quark EMT vertex used in the direct GFF calculation~\cite{Wang:2024abv} onto the light front direction gives
\begin{equation}
\begin{aligned}
  n_{\mu}n_{\nu} \Lambda_q^{\mu\nu}(P_C,\,q)
  &=
  P_C^+\slashed{n}\,F_{1,0}^{q,\mathrm{VC}}(t)
  +
  \frac{(q^+)^2}{m_q}F_{2,0}^{q,\mathrm{VC}}(t)
  \\
  &\quad +
  \frac{P_C^+}{2m_q}i\sigma^{nq}F_{4,0}^{q,\mathrm{VC}}(t).
  \label{eq:gff_meson_cloud}
\end{aligned}
\end{equation}
In the GPD loop of Fig.~\ref{fig:feynman}, the momentum $P_C$ entering this vertex is identified with the loop momentum $k$.

After the light front projection, the first and third terms have the same Dirac and Pauli structures as the dressed electromagnetic current, namely $\slashed{n}$ and $i \sigma^{nq}$.
Their coefficients can therefore be associated with the pairs $F_1^{q,\mathrm{VC}} \leftrightarrow F_{1,0}^{q,\mathrm{VC}}$ and $F_2^{q,\mathrm{VC}} \leftrightarrow F_{4,0}^{q,\mathrm{VC}}$, respectively.
For a bare constituent quark, the charge and mass are carried by the same matter distribution.
Because the bare core weight is large in the present dressed quark current, we assume that the normalized profiles of $F_1^{q,\mathrm{VC}}(t)$ and $F_{1,0}^{q,\mathrm{VC}}(t)$ are similar.
However, this $F_{2,0}^{q,\mathrm{VC}}$ term is absent in our GFFs obtained by taking the Mellin moment with the corrected vertex.
Since this term provides the dominant contribution to the D-term correction~\cite{Wang:2024abv}, the present vertex correction is not reliable for the D-term.
We therefore do not show the corresponding numerical result.
In this approximation, the light cone momentum fraction is assigned to the dressed quark as a whole, without resolving its internal quark and antiquark distributions.
The $u$ and $d$ labels in the numerical results refer to these effective dressed quark contributions.

\section{Numerical Results and Discussion}
\label{sec:results}

\subsection{Model Parameters}

The parameter set used in the numerical calculation is listed in Table~\ref{tab:parameters}.
An explicit transition term between the scalar and axial-vector diquarks is not included here, as it was found to play only a minor role in the nucleon form factor calculations in Ref.~\cite{Wang:2024abv}.
\begin{table}[!htbp]
\caption{\footnotesize{Model parameters used in the numerical calculation of the nucleon GPDs.}}
\label{tab:parameters}
\renewcommand{\arraystretch}{1.0}
\begin{ruledtabular}
\begin{tabular}{llc}
Parameter & Symbol & Value \\
\colrule
Nucleon mass & $M$ & $0.938\,\mathrm{GeV}$ \\
Pion mass & $m_\pi$ & $0.140\,\mathrm{GeV}$ \\
Constituent quark mass & $m_q$ & $0.380\,\mathrm{GeV}$ \\
Diquark mass & $m_D$ & $0.700\,\mathrm{GeV}$ \\
Cutoff mass & $m_R$ & $1.000\,\mathrm{GeV}$ \\
Spin-0 normalization & $c_1^2c^2_s$ & $6.79\,\mathrm{GeV}^8$ \\
Spin-1 normalization & $c_1^2c^2_a$ & $2.19\,\mathrm{GeV}^8$ \\
Mixing parameter & $\sin^2\theta$ & $0.5$ \\
Bare core probability & $Z$ & $0.8$ \\
\end{tabular}
\end{ruledtabular}
\end{table}

\subsection{GPDs and Form Factor Consistency}
\label{sec:ff_consistency}

\begin{figure}[!htbp]
\centering
\includegraphics[width=\columnwidth]{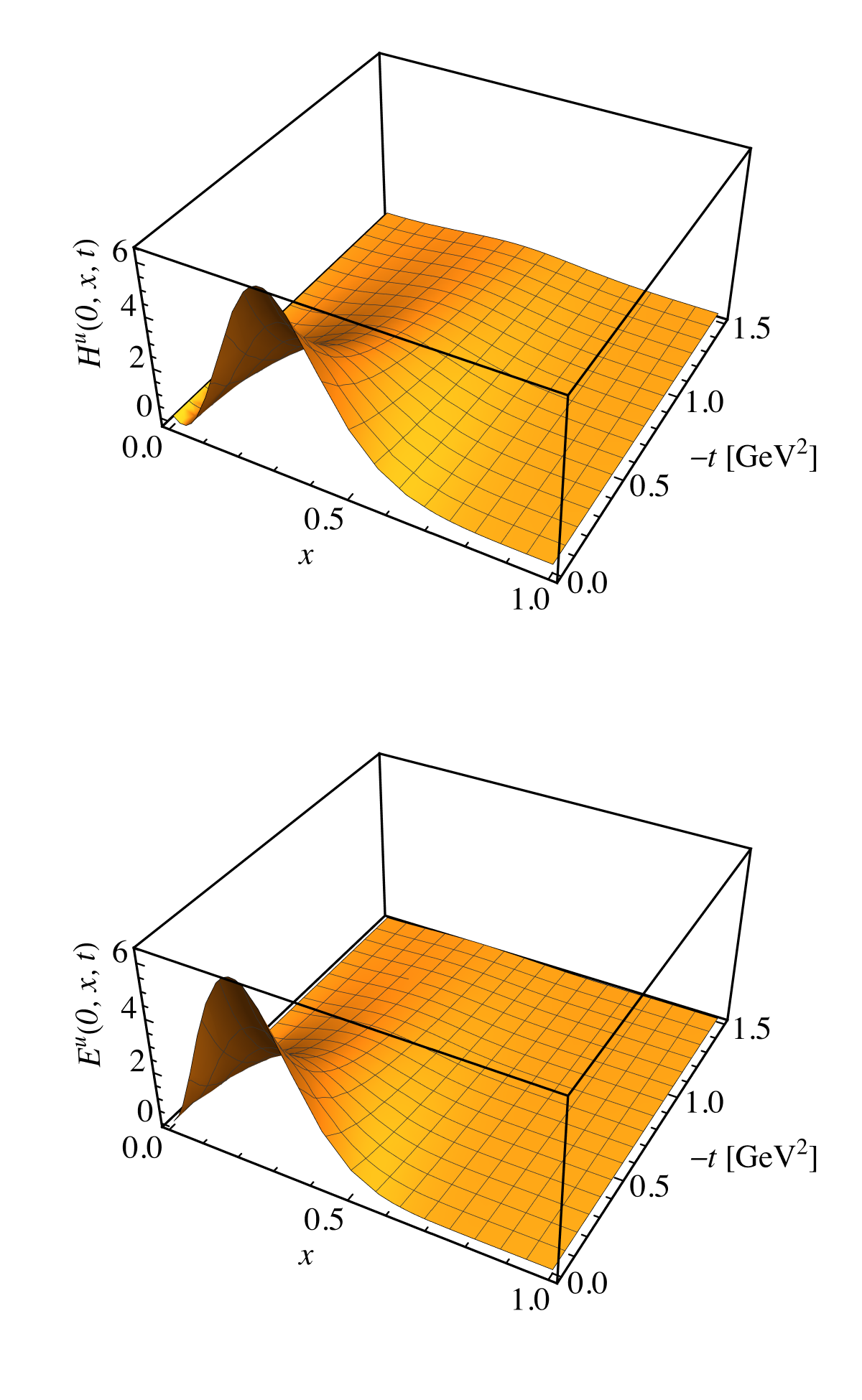}
\caption{\footnotesize{Surfaces of the proton unpolarized GPDs with the vertex correction.
The upper and lower panels show $H^u(0,\,x,\,t)$ and $E^u(0,\,x,\,t)$, respectively.}}
\label{fig:gpd_forward_flavor}
\end{figure}
Figure~\ref{fig:gpd_forward_flavor} shows the proton GPDs $H^u(0,\,x,\,t)$ and $E^u(0,\,x,\,t)$ calculated with the vertex correction.
Further discussion of the GPD results is given in Sec.~\ref{sec:dressed_results}.
At zero skewness, using Eqs.~\eqref{eq:zero sum rules} and~\eqref{eq:first sum rules}, we obtain the electromagnetic form factors $G_{E,M}(t)$ and the mass form factor $A(t)$ from the Mellin moments of the GPDs.
These results agree well with the corresponding direct results of Ref.~\cite{Wang:2024abv}.
This agreement provides a consistency check of the GPD calculation, similar to that used in Ref.~\cite{Fu:2023dea}.

\begin{figure}[!htbp]
\centering
{\includegraphics[width=\columnwidth]{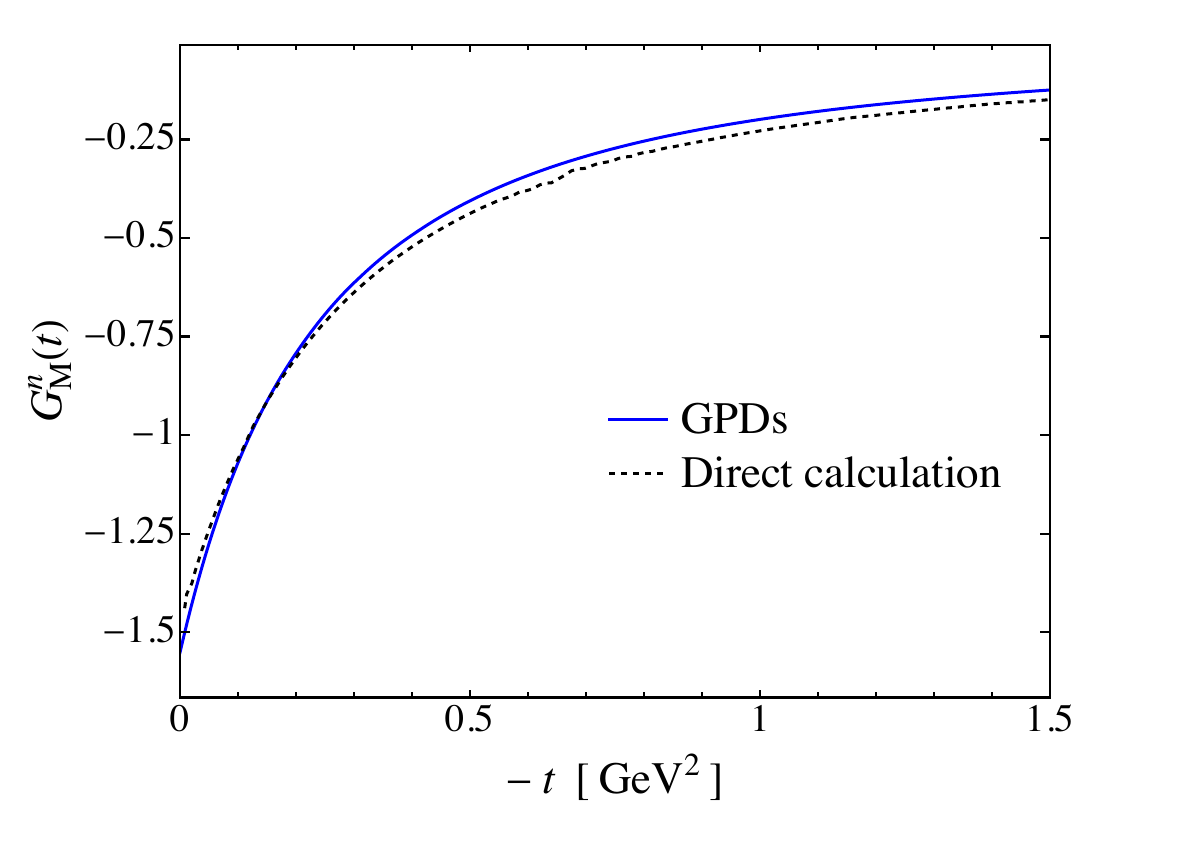}}
\caption{\footnotesize{Neutron magnetic form factor $G_M^n(t)$ obtained from the GPDs with the vertex correction, compared with the direct result of Ref.~\cite{Wang:2024abv}, which includes the pion cloud correction.}}
\label{fig:ff_cloud_consistency}
\end{figure}

Figure~\ref{fig:ff_cloud_consistency} shows the neutron magnetic form factor $G_M^n(t)$ as a representative comparison between the two calculations.
Compared with the calculation in Ref.~\cite{Wang:2024abv}, our calculation omits the longitudinal term of the axial-vector diquark propagator, the momentum dependent term of the axial-vector vertex, and the transitions between scalar and axial-vector diquarks.
These omissions account for the small difference between the two calculations.
In Ref.~\cite{Wang:2024abv}, including pion cloud effects increased the magnitudes of the magnetic form factors and improved their agreement with the experimental data.
We do not repeat the broader discussion of form factors, which has already been given in Ref.~\cite{Wang:2024abv}.

\subsection{Effects of the Vertex Correction}
\label{sec:cloud_comparison}

We determine the model input scale from the forward PDFs.
We evolve the valence distributions according to the LO nonsinglet DGLAP equation directly in $x$ space~\cite{Gribov:1972ri,Altarelli:1977zs,Dokshitzer:1977sg}.
Before evolution, for the proton, the model gives
\begin{align}
  N_u &= \int_0^1 dx\,u(x)=2.00, \\
  N_d &= \int_0^1 dx\,d(x)=1.00, \\
  M_{ud} &= \int_0^1 dx\,x[u(x)+d(x)]=1.00.
\end{align}
These results show that the valence quarks carry all of the nucleon momentum.
This is consistent with a low model input scale~\cite{Scopetta:2002xq}.

\begin{figure}[!htbp]
\centering
{\includegraphics[width=\columnwidth]{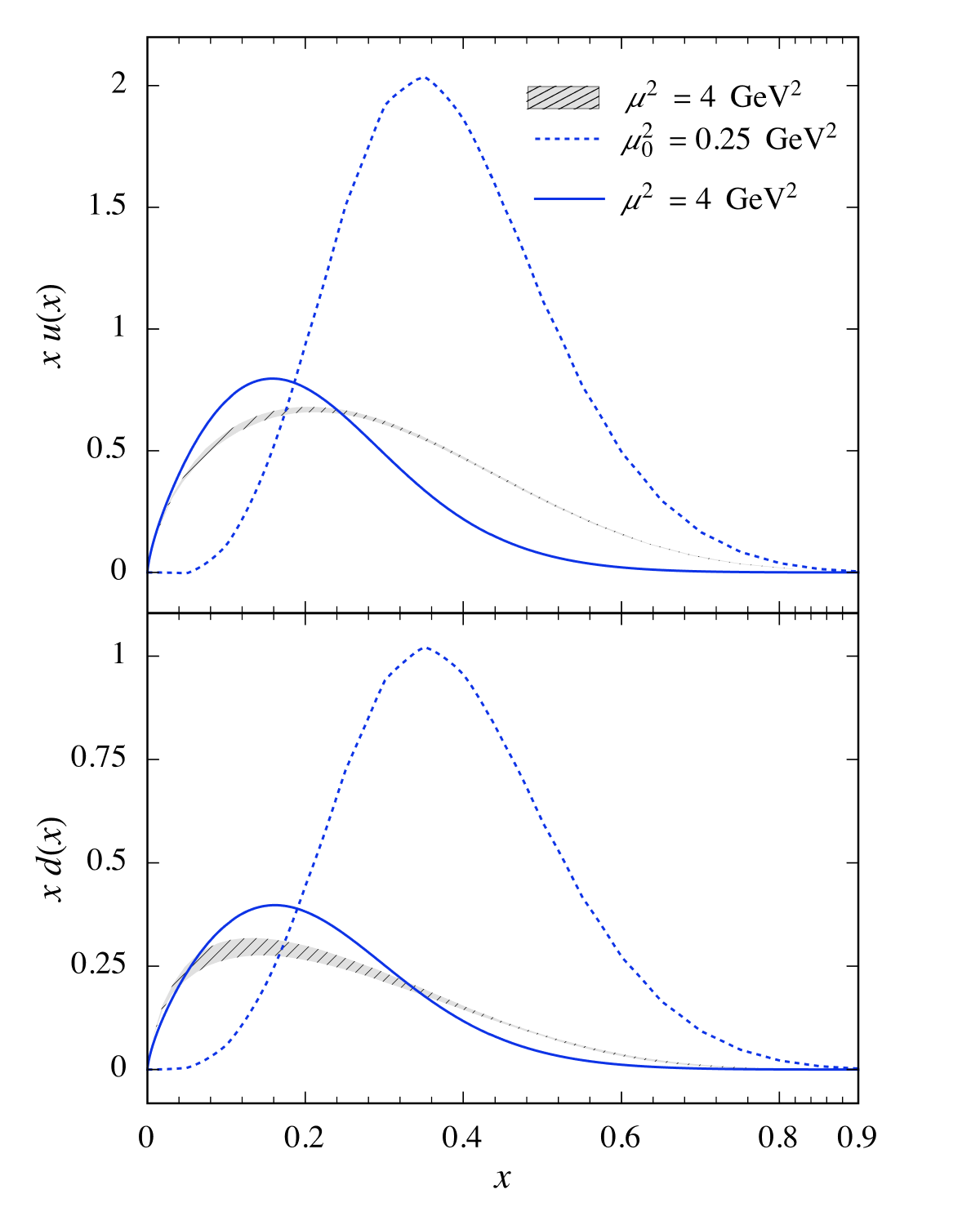}}
\caption{\footnotesize{Proton valence PDFs $xu$ (upper) and $xd$ (lower), obtained from the forward limit $H^q(0,\,x,\,0)$ with the vertex correction.
The dashed and solid curves show the model results at $\mu_0^2=0.25\,\mathrm{GeV}^2$ and after LO evolution to $\mu^2=4\,\mathrm{GeV}^2$, respectively.
The gray hatched band shows the phenomenological PDF range at the latter scale, read off from Fig.~11 of Ref.~\cite{Diehl:2013xca}.}}
\label{fig:pdf_evolution_scale}
\end{figure}
Figure~\ref{fig:pdf_evolution_scale} shows the proton forward PDFs obtained with the vertex correction and evolved at LO to $\mu^2=4\,\mathrm{GeV}^2$, compared with the phenomenological results of Ref.~\cite{Diehl:2013xca}.
We examined model input scales in the range $0.20~\mathrm{GeV}^2 \leq \mu_0^2 \leq 0.30~\mathrm{GeV}^2$.
Guided by the typical low hadronic scales used in constituent quark models~\cite{Scopetta:2002xq,Ossmann:2004bp,Freese:2019eww} and this forward PDF comparison, we choose $\mu_0^2=0.25\,\mathrm{GeV}^2$ as the model input scale.
Under the LO nonsinglet evolution, the valence quark numbers are conserved, while the strength of $xH(0,\,x,\,0)$ is shifted toward smaller $x$.
This indicates that part of the nucleon momentum is redistributed to gluons and sea quarks~\cite{Altarelli:1977zs,Dokshitzer:1977sg}.
The faster decrease of our PDFs at large $x$ may partly reflect the diquark convolution in Eq.~\eqref{eq:xi0_convolution}.
As $x\to1$, the diquark and its active quark must each carry nearly all of the momentum of their respective parent systems, suppressing the contribution from quarks inside the diquark.

\begin{figure}[!htbp]
\centering
\includegraphics[width=\columnwidth]{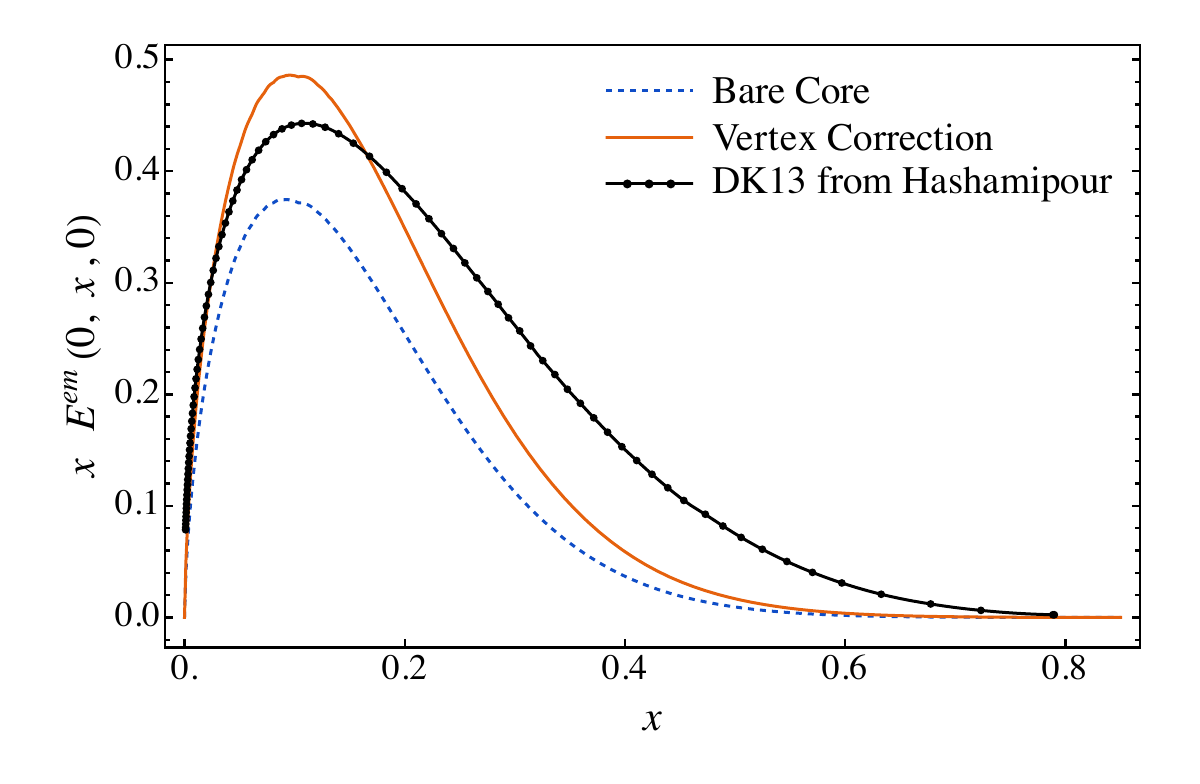}
\caption{\footnotesize{The evolved proton quantity $x E^{em}(0,\,x,\,0)$ at $\mu^2=4\,\mathrm{GeV}^2$ compared with the DK13 parametrization~\cite{Diehl:2013xca}, as shown in Figs.~8 and 9 of Ref.~\cite{Hashamipour:2022noy}.}}
\label{fig:PDF_E_cloud_compare}
\end{figure}
Figure~\ref{fig:PDF_E_cloud_compare} shows the evolved $xE^{em}(0,\,x,\,0)=x (\frac{2}{3}E^u(0,\,x,\,0)-\frac{1}{3}E^d(0,\,x,\,0))$ at $\mu^2=4\,\mathrm{GeV}^2$ with and without the vertex correction.
The zeroth Mellin moment of $E^{em}(\xi, \, x, \, t)$ gives the proton Pauli form factor $F_2^p(t)$.
This common electromagnetic constraint provides a basis for the model comparison.
The DK13 zero skewness parametrization~\cite{Diehl:2013xca}, taken from Figs.~8 and 9 of Ref.~\cite{Hashamipour:2022noy}, is shown for comparison.
The reference $E^u$ and $E^d$ distributions are combined with the same charge weights.
The vertex correction leaves the forward PDFs $H^q(0,\,x,\,0)$ unchanged.
We therefore evolve $E^q$ with and without the correction using the same parameters.
The vertex correction increases the magnitude of $E^{em}$ and brings the distribution closer to the parametrization.
For the zeroth moment $\int_0^1 dx\,E^{em}(0,\,x,\,0)$, we obtain $1.18$ without the vertex correction and $1.53$ with it.
The latter is closer to the empirical proton anomalous magnetic moment, $F_2^p(0)\simeq1.79$.
The improved agreement for both the $E^{em}$ distribution and its zeroth moment provides a phenomenological check of the effective vertex correction.
There is a similar correction effect for the neutron.

\begin{figure*}[!t]
\centering
\includegraphics[width=1.8\columnwidth]{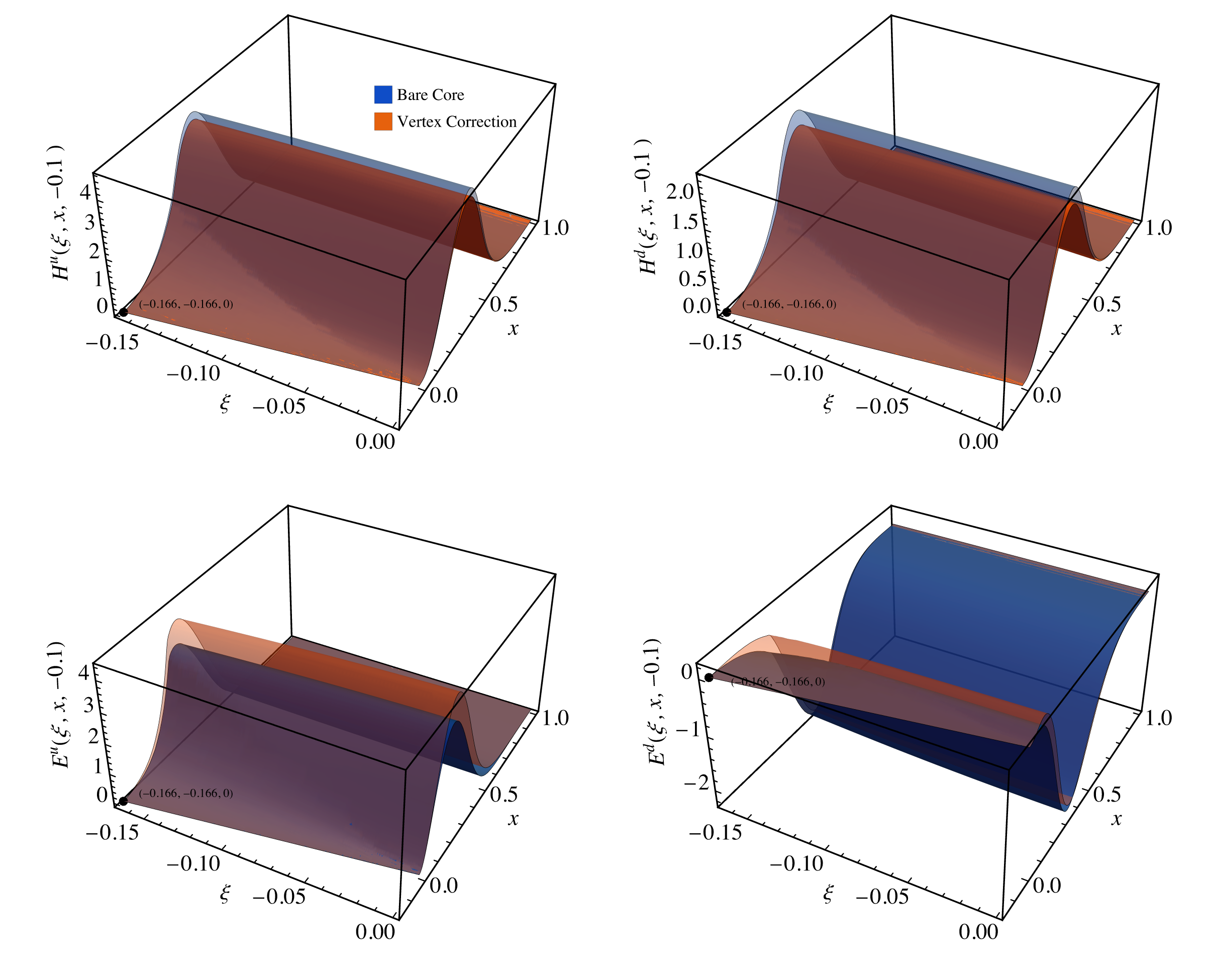}
\caption{\footnotesize{Comparison of the proton $u$ and $d$ quark GPDs at $t=-0.1\,\mathrm{GeV}^2$.
The upper and lower panels show $H^q$ and $E^q$, respectively, with $q=u$ on the left and $q=d$ on the right.
The blue and orange surfaces denote the results without and with the vertex correction, respectively.
The accessible region is limited by $|\xi| \le \sqrt{-t/(4M^2-t)}$, with the lower support boundary $x=-|\xi|$.
The unit of $t$ is omitted in the axis labels.}}
\label{fig:gpd_cloud_compare}
\end{figure*}
Figure~\ref{fig:gpd_cloud_compare} shows the proton GPDs $H^{u,d}(\xi,\,x,\,t)$ and $E^{u,d}(\xi,\,x,\,t)$ with and without the vertex correction at $t=-0.1\,\mathrm{GeV}^2$.
The correction suppresses $H^q$, while increasing the values of $E^q$.
The additional Pauli term in Eq.~\eqref{eq:dressed_vertex} contributes to this increase.
The increase in the values of $E^{u,d}(\xi,\,x,\,t)$ is consistent with the known role of pion cloud effects in the Pauli and magnetic form factors~\cite{Pasquini:2006dv,Pasquini:2006ib,Cloet:2014rja,Wang:2024abv}.
The pion loop corrections involve $u\pi^0$ and $d\pi^+$ intermediate states for a dressed $u$ quark, and $d\pi^0$ and $u\pi^-$ states for a dressed $d$ quark~\cite{Cloet:2014rja}.
The different charge weights of the quark and pion contributions lead to flavor dependent vertex corrections, producing the distinct changes in the $u$ and $d$ quark GPDs shown in Fig.~\ref{fig:gpd_cloud_compare}.

\begin{figure}[!htbp]
\centering
\includegraphics[width=\columnwidth]{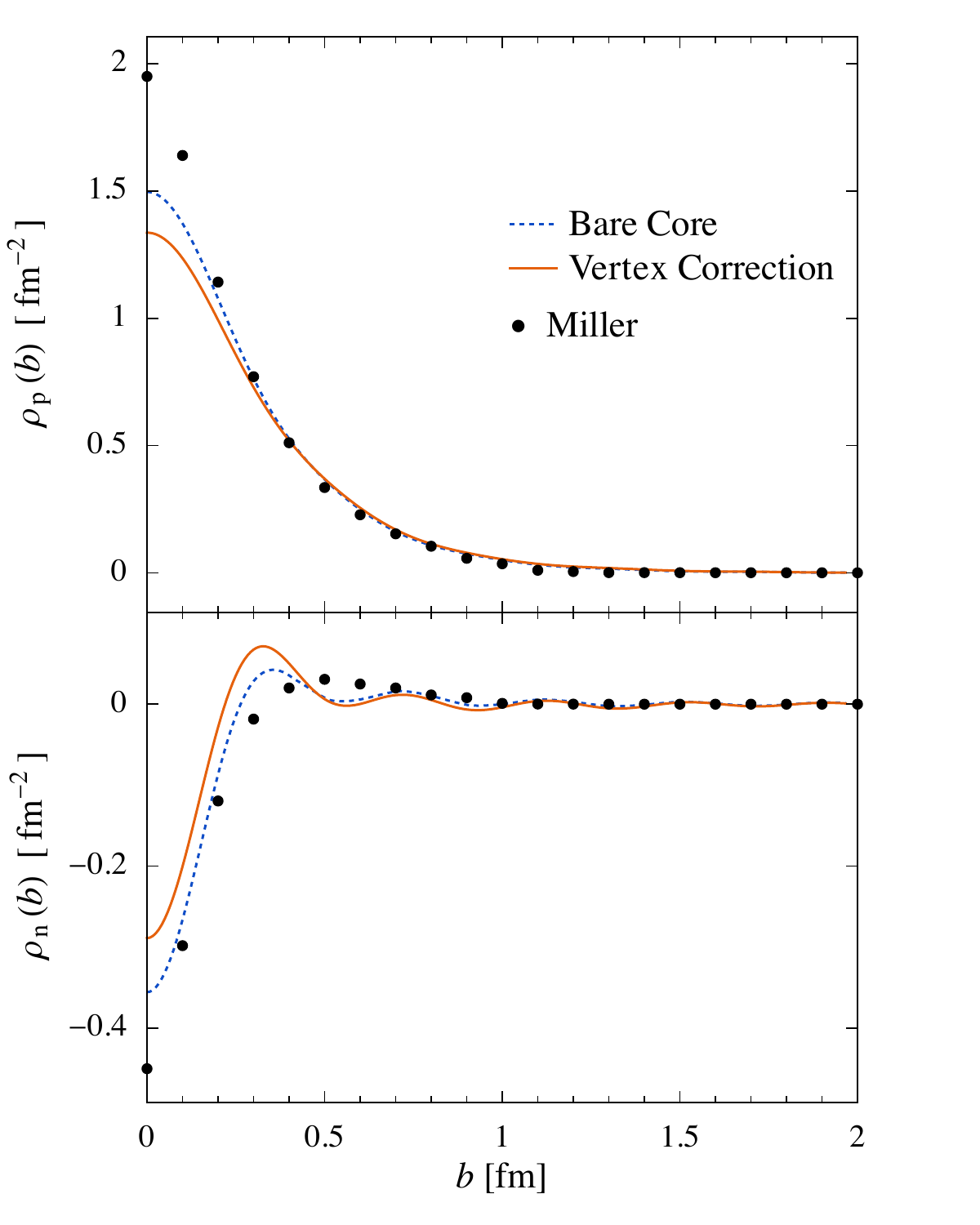}
\caption{\footnotesize{Dirac transverse charge densities of the proton (upper) and neutron (lower) compared with the reference results~\cite{Miller:2007uy}.}}
\label{fig:charge_density_cloud}
\end{figure}
Figure~\ref{fig:charge_density_cloud} shows the transverse charge densities of the proton and neutron, obtained by integrating the corresponding IPDs $q_N(x,\,b)$ over $x$ according to Eq.~\eqref{eq:transverse_charge_density}.
The Dirac form factor $F_1^N(t)$ is Lorentz invariant~\cite{Perdrisat:2006hj,Lorce:2024ipy}.
Thus, the form factor extracted from our GPDs in the Breit frame can be used in this expression.
The results of Ref.~\cite{Miller:2007uy}, obtained from empirical form factors, are also shown for comparison.
The proton exhibits a positive central peak that decreases with increasing $b$, while the neutron shows a negative core followed by a positive peak at intermediate $b$, as in the reference analysis~\cite{Miller:2007uy}.
Both densities approach zero at large $b$.
For both nucleons, the vertex correction reduces the magnitude of the central transverse charge density while preserving the charge normalization, $2\pi\int_0^\infty db\,b\,\rho_N(b)=F_1^N(0)$.
Our results show a positive tail for the proton and a negative tail for the neutron, consistent with the empirical analysis of Ref.~\cite{Miller:2007uy}.
With the charge normalization preserved, the reduction in the magnitude of the central contribution must be compensated at larger transverse distances.
This indicates a redistribution of strength from the central region toward the periphery.

The residual oscillations in the transverse charge densities arise from the sharp cutoff of the Fourier--Bessel integral in Eq.~\eqref{eq:transverse_charge_density}.
We consider a smooth form factor that decreases faster than $1/Q^2$ at large $Q$, consistent with the asymptotic behavior expected from perturbative QCD~\cite{Lepage:1980fj}.
The leading oscillation amplitude can then be estimated as
\begin{equation}
  \mathcal{A}_N(b)\simeq
  \frac{\left|F_1^N(-Q_{\max}^2)\right|}{2\pi b}
  \sqrt{\frac{2Q_{\max}}{\pi b}},
  \qquad Q_{\max}b\gg1.
  \label{eq:cutoff_oscillation_amplitude}
\end{equation}
At $b=1.60\,\mathrm{fm}$ and $Q_{\max}^2=10\,\mathrm{GeV}^2$, this estimate gives $\mathcal{A}_p\simeq7.93\times10^{-4}\,\mathrm{fm}^{-2}$ and $\mathcal{A}_n\simeq2.93\times10^{-3}\,\mathrm{fm}^{-2}$ for the results with the vertex correction.
The neutron transverse charge density is small in magnitude in this region, making the cutoff oscillations particularly noticeable.

\subsection{Further Results with the Vertex Correction}
\label{sec:dressed_results}

\subsubsection{Transverse Structure}

\begin{figure}[!htbp]
\centering
\includegraphics[width=\columnwidth]{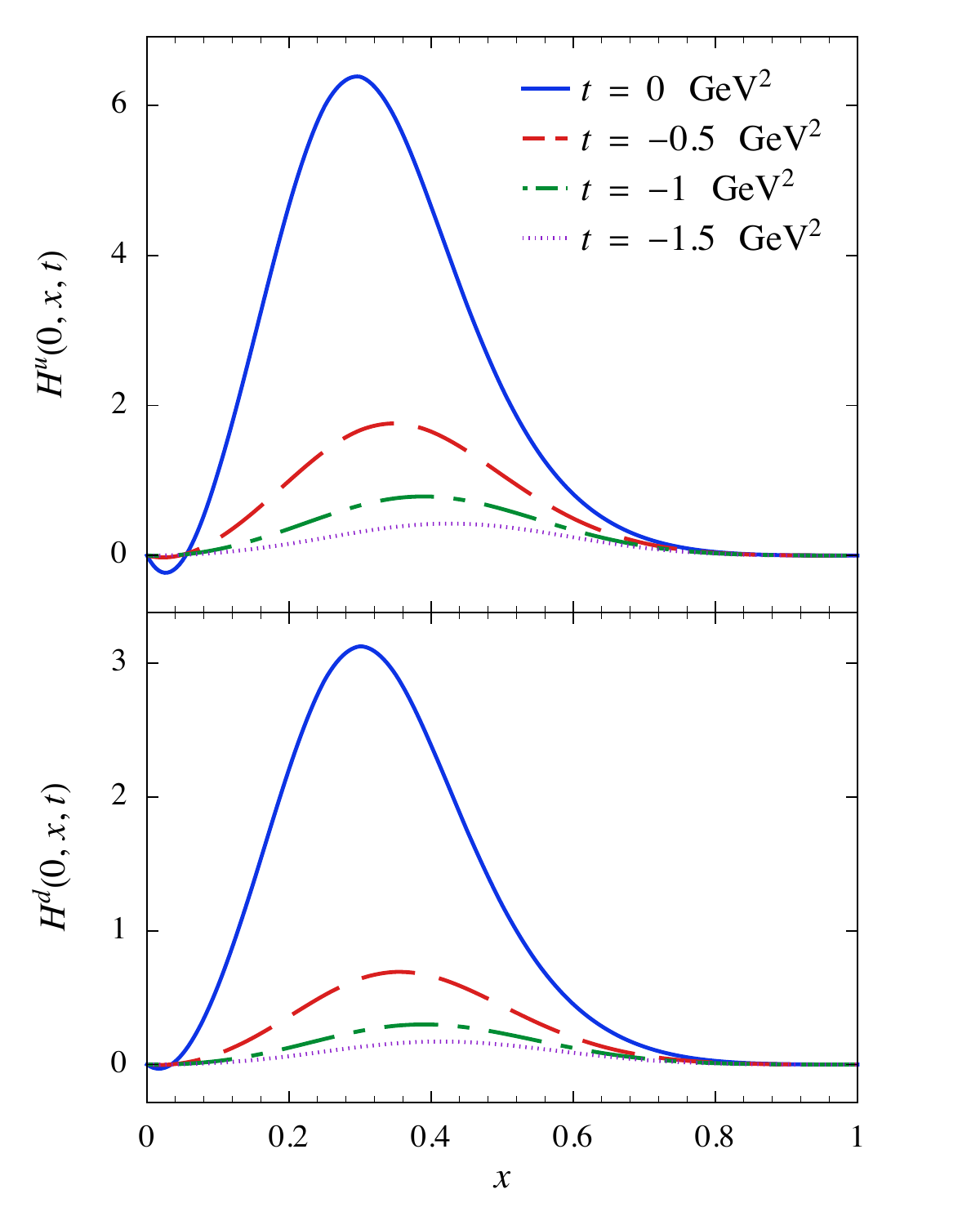}
\caption{\footnotesize{Representative two-dimensional slices of the proton GPD $H^q(0,\,x,\,t)$ with the vertex correction at different $t$.
The upper and lower panels show the $u$ and $d$ quark contributions, respectively.}}
\label{fig:gpd_forward_slices}
\end{figure}
Figure~\ref{fig:gpd_forward_slices} shows two-dimensional slices of the proton GPD $H^q(0,\,x,\,t)$ with the vertex correction for $u$ and $d$ quarks at several values of $t$.
At $t=0$, both distributions peak near $x=\frac{1}{3}$, consistent with the discussion in Ref.~\cite{Fu:2023dea}.
This indicates that, after including the internal quark structure of the diquark, the model still preserves the expected three valence quark momentum sharing pattern, with light quarks of the same mass carrying comparable light cone momentum.
As $|t|$ increases, the larger-$x$ components retain more relative strength, shifting the peak toward larger $x$ in qualitative agreement with Ref.~\cite{Diehl:2004cx}.
At zero skewness, this correlation between $x$ and $t$ is connected to the transverse spatial distribution through the Fourier transform in Eq.~\eqref{eq:ipd_def}.
Larger $|t|$ probes shorter transverse distance scales.
The peak shift thus suggests that larger momentum fractions are relatively more prominent at smaller transverse distances.

\begin{figure}[!htbp]
\centering
\includegraphics[width=\columnwidth]{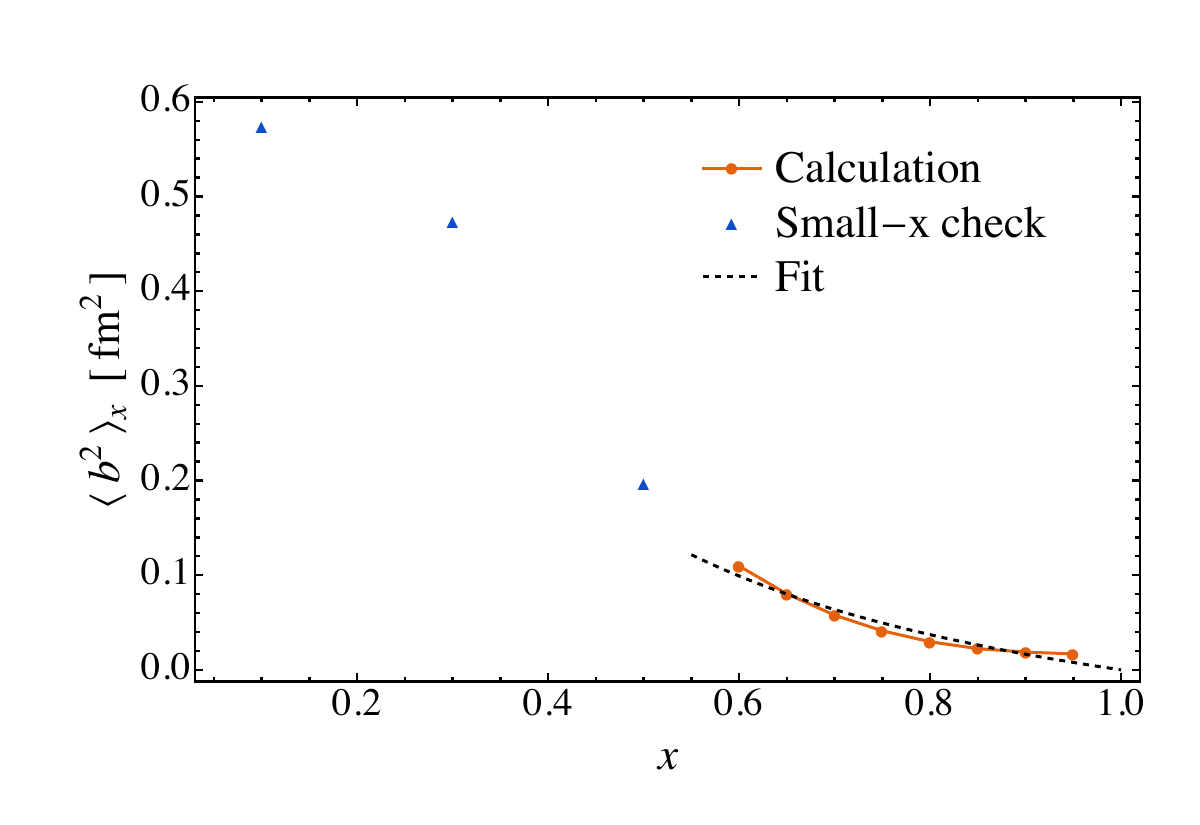}
\caption{\footnotesize{The normalized second transverse moment $\langle \bm b^2\rangle_x$ of the proton IPD with the vertex correction.
The orange points and connecting line show the numerical results at $x\geq 0.6$.
The blue triangles show the additional results at $x=0.1$, $0.3$, and $0.5$, which are not included in the fit.
The black dashed curve is the fit using Eq.~\eqref{eq:ipd_b2_gaussian}.
Here we omit the proton label in $\langle \bm b^2\rangle_{x,p}$ for brevity.}}
\label{fig:ipd_b2_fit}
\end{figure}
Figure~\ref{fig:ipd_b2_fit} shows the calculated $\langle \bm b^2\rangle_x$ for the proton and the fitted curve.
We fit our large $x$ results using the form discussed in Ref.~\cite{Diehl:2004cx}.
This form follows from a model with Gaussian light cone wave functions for Fock states with few partons.
The model provided a reasonable description of quark distributions for $x\gtrsim 0.6$--$0.7$.
The corresponding normalized second transverse moment takes the form
\begin{equation}
  \langle \bm b^2\rangle_x \simeq 2a^2\frac{1-x}{x}.
  \label{eq:ipd_b2_gaussian}
\end{equation}
The fit gives $a=0.27\,\mathrm{fm}$ for the transverse size parameter.
The blue triangles show the results at $x=0.1$, $0.3$, and $0.5$, which are not included in the fit.
These additional points at $x<0.6$ also follow the  same trend: $\langle \bm b^2\rangle_x$ decreases as $x$ increases.
These results suggest a harder longitudinal momentum distribution in the central transverse region of the nucleon than in its periphery.

\subsubsection{ERBL Sensitivity}
\label{sec:erbl_completion}

\begin{figure}[!htbp]
\centering
{\includegraphics[width=\columnwidth]{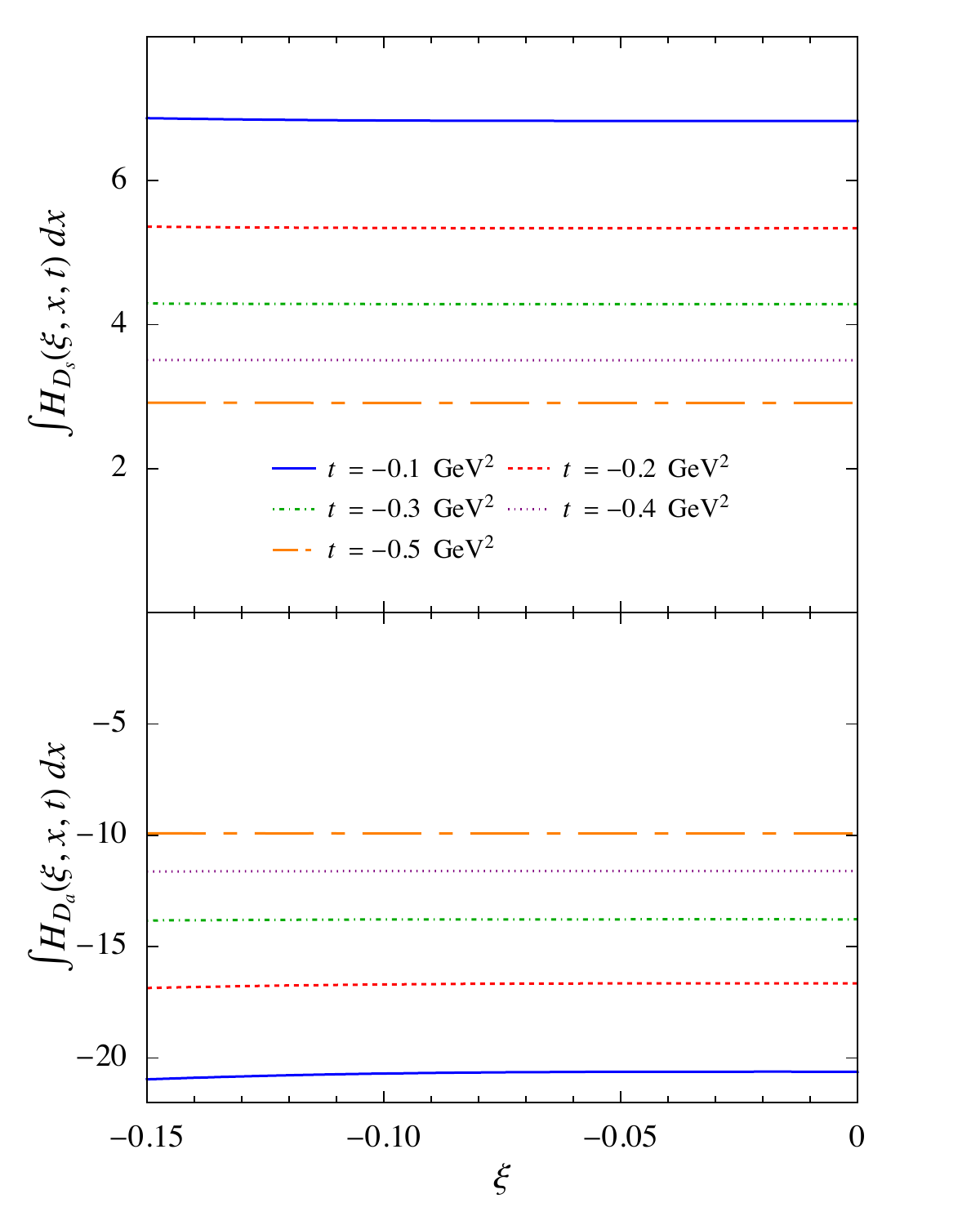}}
\caption{\footnotesize{The zeroth Mellin moment stability check for the proton at finite skewness.
The upper and lower panels show the $\xi$ dependence of $\int dx\,H_{D_{s(a)}}(\xi,\,x,\,t)$ for the spin-0 and spin-1 diquark channels after convolution, respectively, at different values of $t$.
The results are unnormalized.}}
\label{fig:xi_check_diquark_mellin}
\end{figure}
Figure~\ref{fig:xi_check_diquark_mellin} shows the zeroth Mellin moment test of the proton separately for the scalar and axial-vector diquark channels without the GDA contribution.
For a complete construction, the integral of $H$ over $x$ should be independent of $\xi$ at fixed $t$, as required by the sum rule in Eq.~\eqref{eq:zero sum rules}.
We use this property to estimate the numerical effect of the missing GDA contribution in the ERBL region.
At $\xi=0$, the GDA contribution is absent, and the result corresponds to the associated form factor up to the omitted normalization factors.
As $\xi$ changes, the integral of $H$ over $x$ shows only a weak variation.
This indicates that, in the small $|\xi|$ region considered here, neglecting the GDA contribution has only a small effect on the zeroth Mellin moment.

We evolve the nonforward GPDs from the input scale fixed by the forward PDF comparison to the hard scale used in the phenomenological analysis.
The corresponding GPD evolution kernels are applied in $(\xi,\,x)$ space, with the singlet and nonsinglet parts treated according to the DGLAP and ERBL regions~\cite{Diehl:2003ny,Belitsky:2005qn}.
The evolved GPDs are then used to calculate the CFFs according to Eq.~\eqref{eq:cff_definition}.

In our calculation, the sign of $\xi$ is opposite to that used in the cited CFF extraction~\cite{Le:2025swl}, so we compare the physical trend in terms of $|\xi|$~\cite{Belitsky:2001ns,Belitsky:2005qn,Guidal:2013rya}.

\begin{figure}[!htbp]
\centering
\includegraphics[width=0.9\columnwidth]{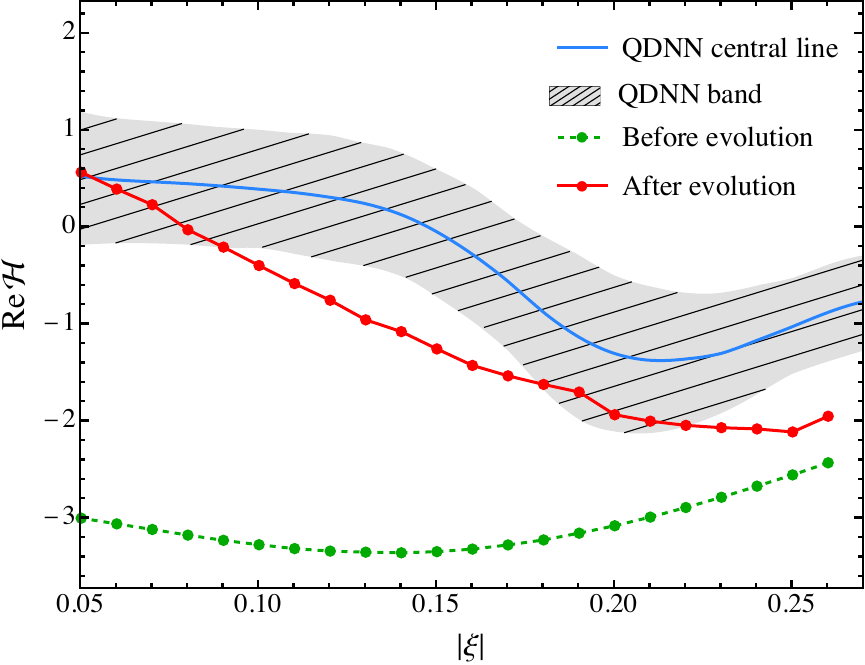}
\caption{\footnotesize{Comparison of the proton $\mathrm{Re}\,\mathcal{H}$ calculated with the vertex correction at $\mu^2=2\,\mathrm{GeV}^2$ and $t=-0.3\,\mathrm{GeV}^2$ with the Quantum Deep Neural Networks (QDNN) extraction.
The blue central line and gray hatched band are taken from Fig.~14(a) of Ref.~\cite{Le:2025swl}; the green dashed curve is the unevolved result, while the red curve is the result after evolution.}}
\label{fig:cff_qdnn_compare_withpion}
\end{figure}
Figure~\ref{fig:cff_qdnn_compare_withpion} shows the proton $\mathrm{Re}\,\mathcal{H}$ at $\mu^2=2\,\mathrm{GeV}^2$ and $t=-0.3\,\mathrm{GeV}^2$, compared with the QDNN extraction shown in Fig.~14(a) of Ref.~\cite{Le:2025swl}.
After LO evolution, the result is comparable in overall magnitude to the QDNN extraction.
It decreases first and then turns upward as $|\xi|$ increases, qualitatively resembling the trend of the QDNN extraction.
Its value remains more negative than the QDNN central line over most of the plotted range.
This residual deviation may reflect the missing GDA contribution in the ERBL region.
The use of LO evolution adds a further approximation.

The larger deviation found for $\mathrm{Re}\,\mathcal{H}$ may reflect its greater sensitivity to the ERBL region compared with the other quantities discussed above~\cite{Belitsky:2001ns,Belitsky:2005qn,Polyakov:1999gs,Chouika:2017dhe}.
The LO hard scattering kernel in Eq.~\eqref{eq:cff_C_definition} has poles at $x=\pm\xi$.
The principal value integral for $\mathrm{Re}\,\mathcal{F}$ is sensitive to the GPD behavior near the crossover lines $x=\pm|\xi|$ and also receives contributions from the ERBL region.

\section{Summary and Outlook}
\label{sec:summary}

In this work, we studied the unpolarized GPDs of the nucleon in a covariant quark--diquark approach.
The internal quark structure of the diquark was included in the nucleon GPDs through a convolution formula.
Guided by the previous form factor study, we introduced an effective vertex correction to include pion cloud effects at the GPD level.

At the model input scale, the forward limit gave $N_u=2.00$, $N_d=1.00$, and $M_{ud}=1.00$, indicating an input with only valence quarks.
The peak positions near $x=\frac{1}{3}$ indicated that the three valence quarks with the same constituent mass shared the light cone momentum approximately equally.
As $|t|$ increased, the peak moved toward larger $x$.
Moreover, the normalized second transverse moment decreased as $x$ increased.
This indicated a relative enrichment of larger-$x$ components in the central transverse region of the nucleon, suggesting a harder longitudinal momentum distribution there than in the periphery.

With the vertex correction, the electromagnetic form factors $G_{E,M}(t)$ and the mass form factor $A(t)$ obtained from the Mellin moments of the GPDs agreed well with those obtained from direct calculations.
The remaining small differences arose from three omissions in the present calculation.
These were the longitudinal term $k^\alpha k^\beta/m_D^2$ of the axial-vector diquark propagator, the momentum dependent term of the axial-vector vertex, and the transitions between scalar and axial-vector diquarks.
The correction preserved the forward PDFs and the valence quark numbers, while improving the magnetic observables associated with the Pauli type distribution $E(\xi,\,x,\,t)$.
The additional Pauli term introduced by the vertex correction brought the evolved $E^{em}$ distribution closer to the DK13 parametrization.
The vertex correction mainly redistributed the transverse strength, shifting it from the central region toward larger transverse distances.
This behavior was consistent with the long range character of the pion cloud contribution.

For the real part of the proton CFF, $\mathrm{Re}\,\mathcal{H}(\xi,\,t)$, the evolved result had a reasonable magnitude and followed the main trend of the QDNN extraction.
The comparison suggested that $\mathrm{Re}\,\mathcal{H}$ might be more sensitive to the ERBL region than the other physical quantities calculated in this work.

Several limitations remain and suggest natural extensions.
First, the GDA contribution is not included in the present truncated convolution.
A future extension should construct this contribution explicitly and include it in the numerical implementation of Eq.~\eqref{eq:full_convolution}.
Second, the present vertex correction is a phenomenological approximation to the pion cloud effect.
A more complete treatment should calculate the GPDs of a quark dressed by the pion cloud from the pion--quark diagrams and fold them into the nucleon GPDs through the corresponding light cone convolution.
This would resolve the quark and antiquark flavor distributions inside the dressed quarks and allow direct comparisons with phenomenological GPDs for individual flavors.
Addressing these limitations would enable a consistent calculation of higher Mellin moments with pion cloud corrections.
It would also allow a more reliable study of quantities sensitive to the ERBL region, especially the D-term and the real parts of the CFFs.

\begin{acknowledgments}

This work is supported by the National Natural Science Foundation of China under Grant Nos.~11975245, 12375142, and 12447121, and by the Gansu Province Postdoctor Foundation.

\end{acknowledgments}

\bibliography{reference}

\end{document}